\documentclass[prl,twocolumn,superscriptaddress,preprintnumbers,amsmath,amssymb,floatfix,nofootinbib]{revtex4}

\usepackage{graphicx}
\usepackage{amsmath}    
\usepackage[normalem]{ulem}
\usepackage{bm}

\usepackage[utf8x]{inputenc}
\usepackage{multirow}
\usepackage{subfigure}
\usepackage{booktabs}
\usepackage{epsfig,graphics}
\usepackage{amsfonts,amssymb,amsmath}            
\usepackage{amsthm}

\newcommand{\nc}{\newcommand}

\nc{\be}{\begin{equation}}

\nc{\ee}{\end{equation}}

\nc{\bea}{\begin{eqnarray}}

\nc{\eea}{\end{eqnarray}}

\nc{\xx}{\nonumber\\}

\nc{\ct}{\cite}

\nc{\la}{\label}

\nc{\eq}[1]{(\ref{#1})}

\def\ajou#1&#2(#3){\ \sl#1\bf#2\rm(19#3)}

\def\[{\left [}
\def\]{\right ]}

\theoremstyle{plain}

\theoremstyle{definition}

\begin{document}
\title{Topology Change of Spacetime and Resolution of Spacetime Singularity \\ in Emergent Gravity}

\author{Sunggeun Lee}
\affiliation{Department of Physics, Sogang University, Seoul 121-741, Korea}

\author{Raju Roychowdhury}
\affiliation{Center for Quantum Spacetime, Sogang University, Seoul 121-741, Korea}
\affiliation{Center for Theoretical Physics and Department of Physics \& Astronomy,
Seoul National University, Seoul 151-747, Korea}

\author{Hyun Seok Yang}
\affiliation{Center for Quantum Spacetime, Sogang University, Seoul 121-741, Korea}

\date{\today}

\begin{abstract}

Emergent gravity is based on the Darboux theorem or the Moser lemma in symplectic geometry
stating that the electromagnetic force can always be eliminated by a local coordinate transformation
as far as U(1) gauge theory is defined on a spacetime with symplectic structure. In this approach,
the spacetime geometry is defined by U(1) gauge fields on noncommutative (NC) spacetime.
Accordingly the topology of spacetime is determined by the topology of NC U(1) gauge fields.
We show that the topology change of spacetime is ample in emergent gravity and the subsequent resolution
of spacetime singularity is possible in NC spacetime. Therefore the emergent gravity approach provides
a well-defined mechanism for the topology change of spacetime which does not suffer any spacetime
singularity in sharp contrast to general relativity.

\end{abstract}


\maketitle

The general theory of relativity predicts the existence of spacetime singularity at the center of
black holes and the very beginning of our universe. The singularity theorem \cite{hawking-ellis} debunks
that classical general relativity cannot be an ultimate theory of space and time.
In order to avoid the spacetime singularities, one would have to resort to a viable quantum theory
of gravity which requires to consider fluctuations not only in geometry but also in topology.
The topology of spacetime enters general relativity through the fundamental assumption that spacetime
is organized as a (pseudo-)Riemannian manifold. But it was shown \cite{tipler} that generic topology
changing spacetimes are singular and so topology change does not seem to be allowed in classical
general relativity. So far this issue has been discussed largely in the context of Euclidean
quantum gravity \cite{eqg-book}  which is hard to justify from first principles and also difficult
to do reliable calculations since the Euclidean Einstein action is unbounded from below.

But a folklore in quantum gravity is that the description of spacetime using commutative coordinates
is not valid below a certain fundamental scale, e.g., the Planck length and beyond that scale
the spacetime has a noncommutative (NC) structure leading to a resolution of spacetime singularities.
Recently string theory has been successful to resolve certain types of geometrical singularities using
the fuzziness of geometry on string length scales. See, for example, a recent review \cite{emil}.
The reason that a vanishing cycle can be nonsingular in string theory is that strings can sense
not only the volume of the cycle but also the flux of $B$ fields. (See section 3 in \cite{emil}
for this feature and section 4 for the significance of NC geometry for the topology change
and singularity resolution.) Since the emergent gravity is also based on a gauge theory
with $B$-field backgrounds \cite{hsyang}, viz., NC spacetime, it will be interesting to see
whether a similar physics to string theory can arise in the emergent gravity approach.

The basic picture of emergent gravity in \cite{hsyang} is that gravity and spacetime are collective
manifestations of U(1) gauge fields on a NC spacetime. In this approach,
the spacetime geometry is defined by U(1) gauge fields on NC spacetime. Accordingly the topology of
spacetime is determined by the topology of NC U(1) gauge fields. As is now well-known, the topology of
NC U(1) gauge fields is nontrivial and rich \cite{u1-topology} and
NC U(1) instantons \cite{ns-inst} represent the pith of their nontrivial topology.
Recently we illustrated in \cite{lry2} how the nontrivial topology of U(1) gauge fields
faithfully appears in the emergent gravity description. In this paper
we will show that the topology change of spacetime is ample in emergent gravity and
the subsequent resolution of spacetime singularity is possible in NC spacetime.
Here we will present only the main ideas and results attributing further details to \cite{lry2,lry3}.

Emergent gravity is based on the Darboux theorem or the Moser lemma in symplectic geometry \cite{sg-book}
stating that the electromagnetic force can always be eliminated by a local coordinate transformation
as far as U(1) gauge theory is defined on a symplectic manifold $(M, B)$.
Let us introduce dynamical gauge fields $A_\mu (x)$ fluctuating around the background $B = dA^{(0)}$.
The resulting field strength is then given by $\mathcal{F} = B + F$ where $F = dA$.
One may introduce local coordinates $x^a, \; a = 1, \cdots, 4$, on a local chart $U \subset M$
where the symplectic structure ${\mathcal F}$ is represented by
\begin{equation}\label{symp-omega}
    {\mathcal F} = \frac{1}{2}\Big( B_{ab} + F_{ab}(x) \Big) dx^a \wedge
    dx^b.
\end{equation}
According to the Moser lemma in symplectic geometry, one can always find a local coordinate transformation
$\phi: x \mapsto y=y(x)$ to eliminate the electromagnetic force $F$ in the symplectic structure
${\mathcal F}$ on $U \subset M$ such that
\begin{equation}\label{darboux-frame}
    {\mathcal F}|_U = \frac{1}{2} B_{\mu\nu} dy^\mu \wedge
    dy^\nu.
\end{equation}
One can solve the condition (\ref{darboux-frame}) by assuming the coordinate transformation as
\begin{equation}\label{cov-cod}
    x^\mu(y) \equiv y^\mu + \theta^{\mu\nu} \widehat{A}_\nu(y)
\end{equation}
which play the role of covariant (dynamical) coordinates in NC gauge theory. By comparing (\ref{symp-omega}) and (\ref{darboux-frame}), one can state the above Darboux transformation as the relation represented by
\begin{eqnarray} \label{eswmap}
    &&  \widehat{F}_{\mu\nu}(y) = \Bigl(\frac{1}{1 + F\theta} F \Bigr)_{\mu\nu}(x), \\
    \label{measure-sw}
    && d^{4} y = d^{4} x \sqrt{\det(1+ F \theta)}(x),
\end{eqnarray}
where we call $\widehat{A}_\mu(y)$ in Eq. (\ref{cov-cod}) ``symplectic gauge fields" whose
field strength is given by
\begin{equation} \label{symp-f}
 \widehat{F}_{\mu\nu} (y) = \partial_\mu \widehat{A}_\nu (y)
- \partial_\nu \widehat{A}_\mu (y) + \{\widehat{A}_\mu, \widehat{A}_\nu \}_\theta (y).
\end{equation}
It turns out \cite{sw-darboux,hsy-darboux} that the local coordinate transformation
to the Darboux frame is equivalent to the Seiberg-Witten (SW) map defining a spacetime field redefinition
between ordinary and NC gauge fields \cite{ncft-sw}.

NC gauge theory is defined by quantizing the covariant coordinates $x^a(y) \in C^\infty(M)
\mapsto \widehat{x}^a (y) \in \mathcal{A}_\theta$ with the Poisson structure
$B^{-1} \equiv \theta = \frac{1}{2} \theta^{\mu\nu} \partial_\mu \wedge \partial_\nu$ in which
the coordinate generators of $\mathcal{A}_\theta$ are noncommuting with the Heisenberg algebra relation
\begin{equation}\label{nc-space}
    [y^\mu, y^\nu]_\star = i \theta^{\mu\nu}.
\end{equation}
The action of NC U(1) gauge theory is then given by
\begin{equation}\label{nc-action}
   \widehat{S}  = \frac{1}{4} \int d^4 y \widehat{F}_{\mu\nu}  \widehat{F}^{\mu\nu},
\end{equation}
with the NC field strength $ \widehat{F}_{\mu\nu} \in \mathcal{A}_\theta$ defined by
\begin{equation} \label{nc-f}
 \widehat{F}_{\mu\nu}  = \partial_\mu \widehat{A}_\nu
- \partial_\nu \widehat{A}_\mu  -i  [\widehat{A}_\mu,
\widehat{A}_\nu ]_\star.
\end{equation}
Note that the field strength of NC U(1) gauge fields is nonlinear due to the commutator term and so
one can find a nontrivial solution of the self-duality equation defined by
\begin{equation}\label{self-dual-inst}
 \widehat{F}_{\mu\nu} (y) = \pm \frac{1}{2} {\varepsilon_{\mu\nu}}^{\rho\sigma}
 \widehat{F}_{\rho\sigma} (y).
\end{equation}
A solution of the self-duality equation (\ref{self-dual-inst}) is called NC U(1) instantons \cite{ns-inst,kly-jkps02}.

It was shown in \cite{ns-inst} that NC U(1) instantons can be obtained by the Atiyah-Drinfeld-Hitchin-Manin (ADHM) construction whose data are specified by a linear Dirac operator $\mathcal{D}^\dagger$ depending on $\mathbf{z} = (z_1 = y^2 + iy^1, \; z_2 = y^4 + i y^3) \in \mathcal{A}_\theta$:
\begin{equation} \label{nc-adhm}
\mathcal{D}^\dagger (\mathbf{z}) = \left(
                                     \begin{array}{c}
                                       \tau_z \\
                                       \sigma^\dagger_z \\
                                     \end{array}
                                   \right) = \left(
                                     \begin{array}{ccc}
                                              B_2 - z_2 & B_1 - z_1 & I \\
                                              -B_1^\dagger + \overline{z}_1 & B_2^\dagger - \overline{z}_2 & J^\dagger
                                            \end{array} \right)
\end{equation}
with $B_1, B_2 : V \to V, \; I: \mathbb{C} \to V, \; J: V \to \mathbb{C}$ where $V$ is a complex vector
space with dimension $k$. The ADHM construction requires the factorization condition $\mathcal{D}^\dagger
\mathcal{D} = \Delta_k \otimes \mathbf{1}_2$ where $\Delta_k$ is a $k \times k$ matrix and
$\mathbf{1}_2$ is a unit matrix in quaternion space. The factorization condition implies the key equations
\begin{equation}\label{fact-adhm}
  \tau_z \tau^\dagger_z -  \sigma_z^\dagger \sigma_z = 0, \qquad \tau_z \sigma_z = 0
\end{equation}
that can be written as the form $\mu_\mathbb{R} = \eta^3_{\mu\nu} \theta^{\mu\nu}
\equiv 2 \zeta_\mathbb{R}$ and $\mu_\mathbb{C} = \frac{1}{2}(\eta^2_{\mu\nu} + i \eta^1_{\mu\nu})
\theta^{\mu\nu} \equiv  \zeta_\mathbb{C}$ where
\begin{eqnarray} \label{moment-r}
\mu_\mathbb{R} &\equiv& [B_1, B_1^\dagger] + [B_2, B_2^\dagger] + II^\dagger -J^\dagger J,  \\
\label{moment-c}
\mu_\mathbb{C} &\equiv& [B_1, B_2] + IJ.
\end{eqnarray}
In the ADHM construction NC U(1) gauge fields with instanton number $k$ are written in the form
\begin{equation}\label{adhm-a}
    \widehat{A}_\mu (y) = i \psi^\dagger(y) \partial_\mu \psi(y)
\end{equation}
where $\psi(y)$ is a free module over $\mathcal{A}_\theta$ satisfying the equations $\psi^\dagger \psi = 1$
and $\mathcal{D}^\dagger \psi = 0$. It can be shown \cite{ns-inst} that the NC field strength
(\ref{nc-f}) determined by the ADHM gauge field (\ref{adhm-a}) is necessarily self-dual
or anti-self-dual if $\psi$ and $\mathcal{D}$ obey the completeness relation
\begin{equation}\label{adhm-comp}
    \psi \psi^\dagger + \mathcal{D} \frac{1}{\mathcal{D}^\dagger\mathcal{D}} \mathcal{D}^\dagger
    = \mathbf{1}_{2k+1}.
\end{equation}
Therefore the NC generalization of ADHM construction provides the complete set of NC U(1) instantons
with arbitrary topological charge $k$.

Consider a commutative limit $|\theta| \to 0$ where NC gauge fields reduce to symplectic gauge fields
whose field strength is given by Eq. (\ref{symp-f}). Using the SW maps (\ref{eswmap}) and (\ref{measure-sw}),
the action (\ref{nc-action}) in this limit can be written as
\begin{equation}\label{sw-action}
 S = \frac{1}{4} \int d^4 x \sqrt{G} G^{\mu\rho}G^{\sigma\nu} F_{\mu\nu} F_{\rho\sigma},
\end{equation}
where we introduced an effective metric determined by U(1) gauge fields
\begin{equation}\label{eff-metric}
    G_{\mu\nu} \equiv \delta_{\mu\nu} + (F\theta)_{\mu\nu}, \;\;
     (G^{-1})^{\mu\nu} = \Big( \frac{1}{1 + F\theta} \Big)^{\mu\nu} \equiv  G^{\mu\nu}.
\end{equation}
As was argued before, there exists a novel form of the equivalence principle for
electromagnetic force as long as spacetime admits a symplectic structure.
As a result, gravity can emerge from NC U(1) gauge theory as a natural consequence of the equivalence principle
for the electromagnetic force \cite{hsyang}. Hence an interesting question is what kind of four-manifold arises
from a solution of the self-duality equation (\ref{self-dual-inst}). It was proved
in \cite{sty-plb,hsy-prlepl} that the commutative description (\ref{sw-action}) of NC U(1) instantons
via the SW map exactly corresponds to gravitational instantons obeying the half-flat condition
\begin{equation}\label{g-inst}
    R_{abef} = \pm \frac{1}{2}{\varepsilon_{ab}}^{cd} R_{cdef},
\end{equation}
where $R_{abcd}$ is a Riemann curvature tensor. The bottom-up approach of emergent gravity \cite{lry1}
also confirms \cite{lry2} that the Eguchi-Hanson (EH) metric \cite{eh-metric} in Euclidean gravity is
coming from symplectic U(1) gauge fields satisfying the self-duality equation (\ref{self-dual-inst}).

Now we will illustrate why the equivalence between symplectic U(1) instantons and
gravitational instantons proved in \cite{hsy-prlepl} implies the topology change of spacetime and
the NC structure of spacetime is crucial for the resolution of spacetime singularity.
To illuminate the issues, let us consider an explicit solution in general relativity whose metric is assumed
to be of the form
\begin{equation}\label{metric-form}
    ds^2 = A^2(r) (dr^2 + r^2 \sigma_3^2) + B^2(r) r^2(\sigma_1^2 + \sigma_2^2)
\end{equation}
where $r^2 = x_1^2 + \cdots + x_4^2$ and we have introduced a left-invariant coframe
$\{ \sigma^i: i =1,2,3 \}$ for $\mathbb{S}^3$ defined by
\begin{equation}\label{coframe-s3}
    \sigma^i = - \frac{1}{r^2} \eta^i_{\mu\nu} x^\mu dx^\nu.
\end{equation}
The EH metric \cite{eh-metric} takes the form (\ref{metric-form}) with
\begin{equation}\label{metric-ab}
    A^2(r) = \frac{r^2}{\sqrt{r^4 + t^4}} = B^{-2}(r).
\end{equation}
After a little algebra using the expression (\ref{coframe-s3}),
the metric (\ref{metric-form}) can be written as the form \cite{lry2}
\begin{equation}\label{gmetric-gen}
g_{\mu\nu}(x) = \frac{1}{2}(A^2 + B^2) \delta_{\mu\nu} - \frac{1}{r^2} (A^2-B^2)
(\eta^3 \overline{\eta}^i)_{\mu\nu} T^i
\end{equation}
where $T^i \; (i=1,2,3)$ are Hopf coordinates defined by the Hopf map $\pi: \mathbb{S}^3 \to \mathbb{S}^2$.
The effective (emergent) metric (\ref{eff-metric}) is related to the gravitational metric
(\ref{gmetric-gen}) by \cite{sty-plb,hsy-prlepl}
\begin{equation}\label{G-g}
    G_{\mu\nu} (x) = \frac{1}{2} \bigl( \delta_{\mu\nu} + g_{\mu\nu}(x) \bigr),
\end{equation}
and so the U(1) field strength in Eq. (\ref{symp-omega}) is given by
\begin{equation}\label{u1f-gen}
F_{\mu\nu}(x) = f_1(r) \eta^3_{\mu\nu} + f_2(r) \overline{\eta}^i_{\mu\nu} T^i
\end{equation}
where
\begin{equation}\label{ab-metric}
    f_1(r) = 1 - \frac{1}{2}(A^2 + B^2), \quad f_2(r) = - \frac{1}{r^2} (A^2-B^2).
\end{equation}

While turning off the dynamical U(1) gauge fields in Eq. (\ref{u1f-gen}), i.e. $A = B = 1$ in
Eq. (\ref{ab-metric}), one can find that the metric (\ref{gmetric-gen}) becomes flat, i.e., $g_{\mu\nu} = \delta_{\mu\nu}$ and recovers the space $\mathbb{R}^4$.
But, if the symplectic gauge fields in Eq. (\ref{u1f-gen}) are developed, the spacetime evolves
to a curved four-manifold with nontrivial topology whose metric is given by Eq. (\ref{gmetric-gen}).
For example, turning on the intanton gauge fields with (\ref{metric-ab}), the resulting spacetime evolves
to the EH space which contains a non-contractible 2-sphere dubbed as the bolt.
Therefore the emergent gravity clearly verifies the topology change of spacetime
due to U(1) instantons \cite{lry2}. The topology change of spacetime can be more clarified by calculating
the topological invariants of U(1) gauge fields given by Eq. (\ref{u1f-gen}) that are equivalent to
the topological invariants of four-manifolds characterized by the Euler characteristic $\chi(M)$ and
the Hirzebruch signature $\tau(M)$. It was shown in \cite{lry2} that, after turning on
the intanton gauge fields with (\ref{metric-ab}), the topological invariants change from $\chi(M) = 1, \;
\tau(M) = 0$ for $\mathbb{R}^4$ to $\chi(M) = 2, \; \tau(M) = - 1$ for the EH space.
We will show later that the topology change of spacetime is actually generic and ample in emergent gravity.
A similar smooth topology change was also illustrated in \cite{madore,shimada} for two-dimensional
NC Riemann surfaces.

The EH space is a regular geometry without any spacetime singularity \cite{eh-metric} and is coming from
the instanton gauge fields defined by Eq. (\ref{u1f-gen}) with (\ref{metric-ab}) \cite{sty-plb}.
Note that the instanton gauge fields are obtained by the SW map (\ref{eswmap}) from NC U(1) instantons obeying
Eq. (\ref{self-dual-inst}) which can be solved by the ADHM construction given by
$\mu_\mathbb{R} = 2 \zeta_\mathbb{R}$ and $\mu_\mathbb{C} =  \zeta_\mathbb{C}$.
The deformation of the hyper-K\"ahler moment maps $\mu_\mathbb{R}$ and $\mu_\mathbb{C}$ is originated from
the NC structure $\theta^{\mu\nu}$ in Eq. (\ref{nc-space}). But the same deformation can be achieved by
modifying the ADHM equations (\ref{fact-adhm}) by \cite{bn-inst}
\begin{equation}\label{mod-adhm}
  \tau_z \tau^\dagger_z -  \sigma_z^\dagger \sigma_z = 2 \zeta_\mathbb{R},
  \qquad \tau_z \sigma_z =  \zeta_\mathbb{C}
\end{equation}
and instead solving the data (\ref{mod-adhm}) on {\it commutative} $\mathbb{C}^2$.
The corresponding U(1) gauge fields (\ref{adhm-a}) defined by the deformed ADHM data (\ref{mod-adhm}) on
commutative $\mathbb{C}^2$ were obtained in \cite{bn-inst} which we call Braden-Nekrasov (BN) instantons.
The result is given by Eq. (\ref{u1f-gen}) with
\begin{equation}\label{bn-ab}
    A^2(r) = \frac{r^2(r^2 + 2t^2)}{(r^2+t^2)^2}, \quad B^2(r) = \frac{r^4 + r^2t^2 + t^4}{r^2(r^2+t^2)}.
\end{equation}
Thus one may wonder what kind of four-manifold arises from the BN instanton. It is obvious from
our construction that the resulting four-manifold is described
by the metric (\ref{metric-form}) with the coefficients (\ref{bn-ab}).
But it was shown \cite{lry2} that the four-manifold determined by the BN instanton exhibits a
spacetime singularity. For example, the Kretschmann scalar $K$ defined by $K = R_{\mu\nu\rho\sigma}R^{\mu\nu\rho\sigma}$ for the metric of the BN instanton is given by
\begin{equation}\label{bn-k}
   \frac{K}{64 t^8} =  \frac{(2r^2 + 3 t^2)^2}{r^4(r^2+ 2t^2)^6} +
  {\rm regular \; terms}
\end{equation}
which blows up at $r=0$ indicating the presence of a spacetime singularity.

It may be emphasized that the Nekrasov-Schwarz (NS) instantons \cite{ns-inst} and
the BN instantons \cite{bn-inst} are obtained by the same ADHM construction defined by
$\mu_\mathbb{R} = 2 \zeta_\mathbb{R}$ and $\mu_\mathbb{C} =  \zeta_\mathbb{C}$.
The only difference is that the former is defined on the NC space (\ref{nc-space}) while
the latter is defined on the {\it commutative} $\mathbb{C}^2$ which causes the ADHM equations
(\ref{mod-adhm}) to deviate from the standard ones (\ref{fact-adhm}).
The direct deformation in Eq. (\ref{mod-adhm}) also causes the completeness relation
(\ref{adhm-comp}) to fail at a finite number of points called ``freckles" \cite{bn-inst}
where spacetime singularities arise.
It turns out \cite{lry2} that the BN instanton also brings about the same kind of topology change
as the NS instanton as was speculated in \cite{bn-inst}.
The topology change due to the BN instantons can positively be supported by
calculating the topological invariants which are given by $\chi(M) = 2, \; \tau(M) = - 1$.
The Euler number $\chi(M) = 2$ stems from the bolt $\mathbb{S}^2$ in the metric
(\ref{metric-form}) with the coefficients (\ref{bn-ab}). It should be compared to $\chi(\mathbb{R}^4) = 1$
for $\mathbb{R}^4$ which is the case of complete turning off of the dynamical gauge fields, i.e. $A=B=1$
in Eq. (\ref{u1f-gen}). But we observed in Eq. (\ref{bn-k}) that the spacetime geometry after
the topology change becomes singular. It is important to recall that the topology change in this case
occurs in commutative spacetime and so the appearance of spacetime singularity
is rather consistent with the theorem \cite{tipler} for the topology change of spacetime in general relativity.
However the topology change due to the NS instantons does not suffer any spacetime singularity \cite{lry2}
because the spacetime geometry after the topology change becomes the EH space that is manifestly
free from any spacetime singularity.

The first order deviation of the quantum or NC multiplication from the classical spacetime is given by
the Poisson bracket of classical observables. Thus the Poisson bracket of classical observables may
be seen as a shadow of noncommutativity in the quantum world. Since a nonperturbative definition
of quantum gravity is still lacking from a direct quantization of general relativity,
one may adopt the NC U(1) gauge theory as a realization of quantum gravity \cite{hsyang}.
Then in the NC gauge theory gravitational physics at a fundamental level is described by NC operators
(or NC fields in spacetime) and conventional geometry and general relativity arise as collective
phenomena in a regime where the relevant observables are approximately commutative
(see, especially, section 4 in  \cite{emil}). In a NC spacetime such as Eq. (\ref{nc-space}),
the proposed uncertainty relation would be $(\delta y)^2 \gtrsim l_{nc}^2 := |\theta|$
instead of Heisenberg's relation $\delta x \delta p \gtrsim \hbar/2$ \cite{uv-ir}.
As a result, one loses the meaning of ``points" in NC spacetime and there exists a sort
of `minimal physical size' in spacetime geometry.
Consequently there are no solutions where the size of the geometry is smaller than the noncommutativity
scale $l_{nc}$. The above ADHM construction for the NS instantons defined in NC spacetime clearly contrasts
with the situation for the BN instantons which are defined in ordinary commutative spacetime
and so cannot prevent freckles from generating the spacetime singularity.

The ADHM analysis also clarifies the reason why the NC structure of
spacetime is essential to realize the topology change of spacetime
free from spacetime singularities. The commutative space is too
rigid to undergo a change in topology whereas the NC space (\ref{nc-space})
is more flexible for the topology change. And, if spacetime geometry
is emergent from NC gauge fields, the resolution of spacetime
singularity may be a natural consequence due to the spacetime
exclusion such as the UV/IR mixing in NC field theory \cite{uv-ir}.
Since the NC spacetime is the crux for emergent gravity
\cite{hsyang}, the topology change of spacetime would thus be ample in
emergent gravity and the resolution of spacetime singularity should be possible in NC spacetime.
Nevertheless, since our result in Eq. (\ref{gmetric-gen}) only shows the final result
after incorporating a back-reaction of NC gauge fields on spacetime geometry,
the detailed mechanism for the topology change or the singularity resolution
is still remained to be investigated. Instead, it will be interesting to explicitly
demonstrate how U(1) gauge fields are well-defined on a blown up space.
The EH space described by the metric (\ref{metric-form}) with the coefficients (\ref{metric-ab})
contains a nontrivial two-cycle $\mathbb{S}^2$ at the origin $(r=0)$ where the metric is degenerate
to the two-dimensional sphere with the metric $t^2(\sigma_1^2 + \sigma_2^2)$.
So we will examine the commutative U(1) gauge fields for the NS instanton on the blown up space
after a topology change. An underlying feature was already explained in \cite{bn-inst}.
The space blown up at the point $\mathbf{0} = (0,0)$ is simply the space $X$ of pairs
$(\mathbf{z}, l)$ where $\mathbf{z}= (z_1, z_2) \in \mathbb{C}^2$ and $l$ is a complex line
passing through $\mathbf{z}$ and the point $\mathbf{0}$. $X$ projects to $\mathbb{C}^2$ via the map
$p(\mathbf{z}, l)= \mathbf{z}$. The fiber consists of a single point except the point $\mathbf{0}$
where the fiber is the space $\mathbb{CP}^1 = \mathbb{S}^2$ of complex lines passing through
the point $\mathbf{0}$.

The commutative U(1) gauge field for the NS instanton was obtained in \cite{ncft-sw}
and in terms of complex coordinates is given by
\begin{equation}\label{u1-complex}
    A = \frac{i}{2} h(r) (z_1 d\overline{z}_1 - \overline{z}_1 dz_1 + z_2 d\overline{z}_2 - \overline{z}_2 dz_2)
\end{equation}
where
\begin{equation}\label{u1-h}
    h(r) = \frac{1}{2} \Bigl(\frac{\sqrt{r^4 + t^4}}{r^2} - 1 \Bigr).
\end{equation}
Let us decompose the U(1) gauge field (\ref{u1-complex}) as $A = A_0 - A_\infty$ where
\begin{eqnarray}\label{dec-u1-0}
 A_0 &=& \frac{i}{2} h_0(r) (z_1 d\overline{z}_1 - \overline{z}_1 dz_1 + z_2 d\overline{z}_2 - \overline{z}_2 dz_2), \\
\label{dec-u1-inf}
  A_\infty &=& \frac{i}{2} h_\infty(r) (z_1 d\overline{z}_1 - \overline{z}_1 dz_1 + z_2 d\overline{z}_2
  - \overline{z}_2 dz_2),
\end{eqnarray}
and
\begin{equation}\label{dec-h}
    h_0(r) = \frac{t^2}{2r^2}, \quad
   h_\infty(r)= \frac{r^2 + t^2 -\sqrt{r^4 + t^4}}{2r^2}.
\end{equation}
The one-form $A_\infty$ is regular everywhere in $\mathbb{R}^4$ but the one-form $A_0$ has a singularity at $r=0$.
Now we will show that the U(1) gauge field $A_0$ is well-defined on the blown up space $X$.
The total space of the blowup  is a union $X = U \cup U_N \cup U_S$ of three coordinate patches.
The local coordinates in the patch $U_N$ are $(u, \lambda)$ such that $z_1 = u, \;
z_2 = u \lambda$ and the coordinates in the patch $U_S$ are $(v, \tau)$ such that $z_1 = v \tau, \;
z_2 = v$. On these patches $\lambda = z_2/z_1$ and $\tau = z_1/z_2$ parameterize the complex lines passing
through the point $\mathbf{0}$. The third patch $U$ has usual coordinates $(z_1, z_2) \neq \mathbf{0}$.

On $U \cap U_N$, we may write
\begin{equation}\label{a-0}
    A_0 = \frac{i}{4} \Bigl( \frac{u d\overline{u} - \overline{u}du}{|u|^2}
    + \frac{\lambda d \overline{\lambda} - \overline{\lambda} d \lambda}{ 1 + |\lambda|^2} \Bigr)
\end{equation}
while, on $U \cap U_S$,
\begin{equation}\label{a-inf}
    A_0 = \frac{i}{4} \Bigl( \frac{v d\overline{v} - \overline{v}dv}{|v|^2}
    + \frac{\tau d \overline{\tau} - \overline{\tau} d \tau}{ 1 + |\tau|^2} \Bigr).
\end{equation}
Let us define regular U(1) gauge fields as
\begin{equation}\label{reg-u1}
    A_{U_N} = \frac{i}{4} \frac{\lambda d \overline{\lambda} - \overline{\lambda} d \lambda}{ 1 + |\lambda|^2},
    \quad A_{U_S} =  \frac{i}{4} \frac{\tau d \overline{\tau} - \overline{\tau} d \tau}{ 1 + |\tau|^2}.
\end{equation}
One can easily show that on the intersection $U \cap U_N$ the one-forms $A_0$ and $A_{U_N}$
are related via a gauge transformation
\begin{equation}\label{gaugetr-0}
  A_0 = A_{U_N} + \frac{1}{2} d \arg u
\end{equation}
and on the intersection $U \cap U_S$ the corresponding gauge transformation is given by
\begin{equation}\label{gaugetr-inf}
  A_0 = A_{U_S} + \frac{1}{2} d \arg v.
\end{equation}
Finally on the intersection $U_N \cap U_S$ the one-forms $A_{U_N}$ and $A_{U_S}$ are
related via the gauge transformation $ \frac{1}{2} d \arg \lambda = - \frac{1}{2} d \arg \tau$, i.e.,
 \begin{equation}\label{gaugetr-0-inf}
  A_{U_N} = A_{U_S} + \frac{1}{2} d \arg \lambda.
\end{equation}
Therefore we have shown that the one-form $A_0$ and so the commutative U(1) gauge fields
in Eq. (\ref{u1-complex}) are well-defined on the blown up space $X$.

It is interesting to notice \cite{bn-inst} that the U(1) gauge fields in Eq. (\ref{u1-complex})
now carry a nontrivial first Chern class on the blown up space $X$. The restriction of $A$
on the exceptional divisor $E$, defined by the equation $u=0$ in $U_N$ and $v=0$ in $U_S$, induces
the field strength $F|_E = F_{U_N} + F_{U_S}$ with
\begin{equation}\label{divisor-f}
  F_{U_N} = \frac{i}{2} \frac{d \lambda \wedge d \overline{\lambda}}{ (1 + |\lambda|^2)^2},  \quad
  F_{U_S} =  \frac{i}{2} \frac{d \tau \wedge d \overline{\tau}}{(1 + |\tau|^2)^2}.
\end{equation}
Thus we get the first Chern class $c_1(L)$ on the line bundle $L$ given by
\begin{equation}\label{1-chern}
    c_1(L) = \frac{1}{2\pi} \int_E F = 1.
\end{equation}
It may be worthwhile to remark that the behavior of the solution (\ref{u1-complex})
in the commutative limit $\theta \to 0$ (which corresponds to $t^2 \to 0$) becomes severely
singular as $h(r) \sim 1/r^4$ \cite{sty-plb} and so the smooth topology change
is ruined in the commutative spacetime as was expected.

Now we will show that the topology change in emergent gravity is generic in the sense that
it is possible to change even the topology of asymptotic geometry.
For example, incorporating generic U(1) gauge fields whose field strength does not vanish
at asymptotic infinity can lead to the change of spacetime geometry from $\mathbb{R}^4$
to $\mathbb{S}^4$ or $\mathbb{CP}^2$ and thus from a noncompact space to a compact space.
We call such cases ``large" topology change
to differentiate with the previous cases called ``small" topology change.
(Caveat: The small topology change does not mean that a global asymptotic geometry does not change either.
Actually the previous cases change the global asymptotic geometry from $\mathbb{R}^4$
to $\mathbb{R}^4/\mathbb{Z}_2$ \cite{lry2}.) In other words, the topology change in emergent
gravity can accompany even the change of the compactness of spacetime geometry \cite{lry3}.

We will consider three cases for an explicit verification:
\begin{eqnarray} \label{three-ab}
\begin{array}{ll}
(a):&  A^2 =1, \quad B^2 = 1 + \frac{t^2}{r^2}, \\
(b):&  A = B = \frac{t^2}{r^2 + t^2}, \\
(c):& A = B^2 = \frac{t^2}{r^2 + t^2}.
\end{array}
\end{eqnarray}
The corresponding field strengths for each case are given by Eq. (\ref{u1f-gen})
with the coefficients determined by Eq. (\ref{ab-metric}).
One can check that, except the case $(b)$, the U(1) field strengths satisfy the Bianchi identity $dF = 0$
and so locally $F = dA$. We will see soon why the case $(b)$ violates the Bianchi identity.
Let us examine the asymptotic behavior of the field strength (\ref{u1f-gen}) for each case:
At $r \to \infty$, $(a): F_{\mu\nu} \to 0, \; (b)\; \& \; (c): F_{\mu\nu} \to \eta^3_{\mu\nu}$.
Hence the gauge fields for the cases $(b)$ and $(c)$ breed further vacuum condensates
$\langle F_{\mu\nu} \rangle_{\mathrm{vac}} = \eta^3_{\mu\nu}$ superposed on the original
background field $B_{\mu\nu}$ and their asymptotic behavior is rather
unique compared to the NS and BN instantons and the case $(a)$.
The gravitational metric for the cases $(a)$-$(c)$ is given by Eq. (\ref{metric-form})
or (\ref{gmetric-gen}). It is easy to identify the corresponding four-manifolds for each case.
The case $(a)$ is the Burns metric \cite{lebrun} on the blow-up of $\mathbb{C}^2$ at the origin
which is a scalar-flat K\"ahler manifold. And the case $(b)$ describes the four-dimensional
sphere $\mathbb{S}^4$ which is a Euclidean de Sitter space with a cosmological constant
$\Lambda = \frac{12}{t^2}$ \cite{egh-report}. Note that $\mathbb{S}^4$ is a locally conformally flat manifold
but is not a K\"ahler manifold. Finally the case $(c)$ is the famous Fubini-Study metric
on $\mathbb{CP}^2$ \cite{egh-report}. $\mathbb{CP}^2$ is a compact K\"ahler manifold with a cosmological
constant $\Lambda = \frac{6}{t^2}$.
(Frankly speaking, we first wrote down the metrics for the three cases in Eq. (\ref{three-ab})
into the form (\ref{gmetric-gen}) and then read off $A(r)$ and $B(r)$ for each case.
Intentionally we reversed the argument to address the issue at hand in a more accessible way.)

It can be shown \cite{lry2,hsy-prlepl} that the K\"ahler condition for the metric (\ref{gmetric-gen})
is equivalent to the Bianchi identity $dF=0$ for the U(1) field strength (\ref{u1f-gen}),
which can be reduced to the form
\begin{equation}\label{k-bianchi}
    \frac{dB^2}{dr} = \frac{2}{r} \bigl(A^2 - B^2).
\end{equation}
Since $\mathbb{S}^4$ is not a K\"ahler manifold, this result defends the reason
why the gauge fields for the case $(b)$ ignore the Bianchi identity.
By turning on generic gauge fields, the spacetime geometry in Eq. (\ref{three-ab}) undergoes a transition
from $\mathbb{R}^4$ to $M$ with a nontrivial topology. The topological invariants after the transition
are exactly the same as those of the four-manifolds in Eq. (\ref{three-ab}) \cite{lry3,egh-report}.
For example, we get $\chi(M) = 2, \; \tau(M) = -1$ for the Burns metric $(a)$.
However, in the course of transition concomitant with the topology and compactness changes of spacetime,
no spacetime singularity arises. This transition is simply described by introducing generic (large)
dynamical gauge fields and this process is completely well-defined in gauge theory.
All these features are very reminiscent of the situations in string theory where the resolution is limited
by the string scale $l_s^2 := 2\pi \alpha'$ and so there is no operational
way using strings to unambiguously determine geometric structures in regimes smaller than
the string scale \cite{emil}.

In summary, we observed that the topology change in commutative spacetime is a singular process and
the spacetime singularity in that case can be resolved in the NC spacetime (\ref{nc-space}) and
the NC structure of spacetime is crucial for the smooth topology change. Our result implies that some
nonsingular gravitational solutions with nontrivial topology can be realized by considering
nontrivial gauge field configurations on a NC spacetime. Furthermore, our result for the compactness change
may have important implications on string theory compactification.
Extra dimensions might be compactified by U(1) gauge fields with a nontrivial vacuum condensate over there
alike the cases $(b)$ and $(c)$. More study in depth would be required \cite{lry3}.

{\it Acknowledgments}  The research of HSY was supported by
Basic Science Research Program through the National Research
Foundation of Korea (NRF) funded by the Ministry of Education,
Science and Technology (2011-0010597).

\end{document}